# Analysis and Performance Comparison of DVB-T and DTMB Systems for Terrestrial Digital TV

Ming Liu, Matthieu Crussière *member IEEE*, Jean-François Hélard *senior IEEE*, Oudomsack Pierre Pasquero
Institute of Electronics and Telecommunications of Rennes (IETR)
Rennes, France
{first name. last name}@insa-rennes.fr

*Abstract*— Orthogonal frequency-division multiplexing (OFDM) is the most popular transmission technology in digital terrestrial broadcasting (DTTB), adopted by many DTTB standards. In this paper, the bit error rate (BER) performance of two DTTB systems, namely cyclic prefix OFDM (CP-OFDM) based DVB-T and time domain synchronous OFDM (TDS-OFDM) based DTMB, is evaluated in different channel conditions. Spectrum utilization and power efficiency are also discussed to demonstrate the transmission overhead of both systems. Simulation results show that the performances of the two systems are much close. Given the same ratio of guard interval (GI), the DVB-T outperforms DTMB in terms of signal to noise ratio (SNR) in Gaussian and Ricean channels, while DTMB behaves better performance in Rayleigh channel in higher code rates and higher orders of constellation thanks to its efficient channel coding and interleaving scheme.

## I. INTRODUCTION

Transition from analog to digital television (DTV) is a trend worldwide. DTV services can be delivered via satellite, cable and terrestrial broadcasting. Due to its flexibility to both stationary and mobile applications, digital terrestrial television broadcasting (DTTB) has attracted more and more interest in recent years. Nowadays, there are three main DTTB standards around the world: Digital Video Broadcasting-Terrestrial (DVB-T) [1] in Europe, the trellis-coded 8-level vestigial side band (8-VSB) modulation system developed by Advanced Television System Committee (ATSC) [2] in North America, the Integrated Services Digital Broadcasting-Terrestrial (ISDB-T) [3] in Japan. Among them, DVB-T plays the most important role. Since first approved in 1997, DVB-T has become the dominant terrestrial broadcasting standard in Europe and is also popular in other continents. By June 2008, DVB-T services have been launched by 33 countries and territories.

After 12 years of developing, the Chinese Digital Terrestrial/Television Multimedia Broadcasting (DTMB) standard [4] was finally ratified in August 2006, and began to be a mandatory national standard in August 2007. DTMB consists of single carrier modulation (C = 1) and multicarrier modulation (C = 3780) which are originated from two former proposals: the single-carrier ADBT-T (Advanced Digital Television Broadcasting-Terrestrial) and the multi-carrier DMB-T (Digital Multimedia/TV Broadcasting-Terrestrial) respectively, providing flexible combinations of working modes for different application scenarios. Because of the enormous TV market in China and the novel signal processing techniques integrated in it, the Chinese DTMB draws great interests from both industries and researchers.

The orthogonal frequency-division multiplexing (OFDM) is definitely the most popular technique adopted by majority of DTTB standards (DVB-T, ISDB-T and DTMB). This is due to its robustness to frequency selective fading. By implementing inverse fast Fourier transform (IFFT) and FFT at transmitter and receiver sides respectively, OFDM transforms a high speed serial data flow to a set of low speed parallel ones at orthogonal flat fading sub-channels.

Traditionally, a cyclic prefix (CP) is inserted between two consecutive OFDM symbols as guard interval (GI). This solution has been chosen for many standards, namely for DVB-T and ISDB-T. The length of the GI is designed to be longer than that of channel memory. By discarding the CP at the receiver, the inter symbol interference (ISI) is then removed from the received signal. With the assistance of CP, the linear convolution between transmitted signal and channel impulse response (CIR) converts into a circular one i.e. the channel convolution effect is turned to be a set of parallel attenuations in the discrete frequency domain. Hence, the equalization in OFDM can be performed by simply multiplying a coefficient on each subcarrier at the receiver. Thus, the equalization complexity of OFDM is significantly low compared with a time domain equalizer.

As the samples for the CP do not convey useful data, several researchers proposed to replace the CP by known pseudo noise (PN) sequences. This becomes the time domain synchronous OFDM (TDS-OFDM [5], also known as pseudo random postfix OFDM, PRP-OFDM [6] and known symbol padding OFDM, KSP-OFDM [7]). Besides serving as GI, the PN sequence can also be exploited to make channel estimation and synchronization in the time domain. Hence, it is not necessary to insert scattered and continual pilots to the OFDM symbols, which increases the spectrum efficiency. Moreover, since the channel estimation can be performed for each OFDM symbol in the time domain, TDS-OFDM can achieve a fast channel acquisition. However, in contrast with the CP-OFDM, the circularity property mentioned above is no longer obtained and specific algorithms have to be processed at the receiver to

This work is supported by the French national project "Mobile TV World".

restore the cyclicity of the signal. In particular, the PN sequence has to be perfectly removed before demodulation, thus leading to a signal referred to as zero padding OFDM (ZP-OFDM) which has been adopted, for instance, in the WiMedia solution for ultra wide band (UWB) context [9]. The resulting signal can then be demodulated using the estimation methods developed for ZP-OFDM, as proposed in [8] for example.

Chinese DTMB standard (multicarrier mode) is somewhat similar to DVB-T in terms of OFDM transmission scheme, signal constellation, and 8 MHz analog bandwidth. Therefore DVB-T is an ideal counterpart of Chinese DTMB standard to analyze the performance and make comparisons. Although there are some measured results presented in [10], [11] in manners of sensitivity and carrier-to-noise ratio (C/N), there is still not so much reference for DTMB as it for DVB-T. So, it is necessary to carry out a survey on DTMB and make a precise comparison between multicarrier mode of DTMB and DVB-T in order to provide a reference for researchers who may be interested in DTMB.

Rest parts of this paper are organized as follows. Section II describes the main features of DVB-T and DTMB systems. A brief discussion on power factor of both systems is also made in this section. Simulation results for both Gaussian channel and multipath channels are presented in Section III. Conclusions are drawn in Section IV.

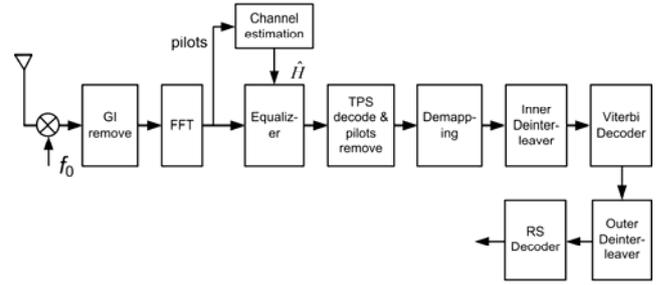

Figure 1. Block diagram of DVB-T receiver

## II. PRESENTATION AND ANALYSIS OF DVB-T AND DTMB SYSTEMS

### A. DVB-T System

DVB-T adopts standard CP-OFDM as transmission scheme. The modulated data symbols are transmitted block-by-block. The $i$th data block $x_M(i)$ is an $M \times 1$ complex vector in the frequency domain whose elements are complex symbols coming from quadrature amplitude (QAM) modulation. After performing $M$-points IFFT, it yields the time domain block:

$$\tilde{x}_M(i) = F_M^H x_M(i) \qquad (1)$$

where $F_M$ is an $M \times M$ FFT matrix with $(m, k)$th entry $M^{-(1/2)}\exp\{-j2\pi nk/M\}$, $(.)^H$ denotes Hermitian transposition, the subscript $M$ indicates its size of either an $M \times 1$ vector or an $M \times M$ matrix and the tilde (˜) denotes time domain variables. Then a CP of length $D$ is inserted between two consecutive blocks. Let $P = M + D$ be the total length of transmitted symbols per block, and let $I_{cp} = [I_c, I_M]^T$ be the $P \times M$ matrix representing the CP appending, where $I_M$ stands for the size $M$ identity matrix and $I_c$ is the $M \times D$ matrix corresponding to the $D$ last columns of $I_M$. The $i$th block of symbols to be transmitted can be expressed as

$$\tilde{x}_{cp}(i) = I_{cp}\tilde{x}_M(i) = I_{cp}F_M^H x_M(i). \qquad (2)$$

The multipath channel can be modeled as an $L$th order FIR filter with impulse response $[h_0, \ldots, h_{L-1}]$. Let $H_{ISI}$ and $H_{IBI}$ be the intra and inter block interference. $H_{ISI}$ and $H_{IBI}$ are $P \times P$ Toeplitz lower and upper triangular matrices with the first column $[h_0, \ldots, h_{L-1}, 0, \ldots, 0]^T$ and first row $[0, \ldots, 0, h_{L-1}, \ldots, h_1]$ respectively. The received $i$th block is:

$$\tilde{r}_p(i) = H_{ISI}\tilde{x}_{cp}(i) + H_{IBI}\tilde{x}_{cp}(i-1) + \tilde{n}_p(i). \qquad (3)$$

Because OFDM system satisfies $D \geq L$, the $H_{IBI}$ can be eliminated by removing the CP. And due to the circular structure of the CP-OFDM, $H_{ISI}$ turns to be an $M \times M$ circulant matrix $H_{circ}$ with the first row $[h_0, 0, \ldots, h_L, \ldots, h_1]$. By the property that circulant matrix is diagonal in Fourier basis, after FFT, the received block becomes:

$$\begin{aligned}r_M(i) &= F_M H_{circ} \tilde{x}_{cp}(i) + F_M \tilde{n}_M(i) \\ &= F_M H_{circ} F_M^H x_M(i) + F_M \tilde{n}_M(i) \\ &= \sqrt{M}\,diag(F_M h_M) x_M(i) + n_M(i)\end{aligned} \qquad (4)$$

where $diag(.)$ denotes a diagonal matrix with elements given by the vector argument, $F_M h_M$ is the frequency response of the multipath channel. Thus, the transmitted signal $x_M$ can be easily recovered from $r_M$ by dividing a corresponding fading factor in the frequency domain.

Fig. 1 presents the block diagram of DVB-T receiver. The GI is first removed from the received symbols. After FFT, pilots are extracted and channel estimation can be made based on them. Then, signals are equalized using the estimated channel frequency response in the frequency domain. The equalized data symbols are then converted to binary bits by demapper. Finally, the erroneous bits are corrected by channel coding combined with interleaving.

Table I gives the key parameters of DVB-T. There are four choices for the GI, providing a guard duration ranging from 7μs to 56μs in 2K mode and 28μs to 224μs in 8K mode. Three constellations can be used with options of hierarchical modes. Both bitwise and symbol interleaving are performed to avoid long sequences of severely corrupted bits feeding to the inner decoder of receiver, which can effectively improve the error correction ability of channel coding in presence of the frequency selective fading. The channel coding consists of Reed-Solomon RS (204, 118, t=8) and punctured convolutional code with code rate 1/2, 2/3, 3/4, 5/6 and 7/8. Between outer and inner coding, a convolutional interleaver with a maximum delay of 2244 bytes is adopted. That means the data is convolutionally interleaved to spread burst errors at the output of the inner decoder over several OFDM blocks, while the bitwise and symbol interleaving are made within one OFDM block.

TABLE I. PARAMETERS OF DVB-T

| Nb. of active subcarriers | 1705 (2K mode), 6817 (8K mode) | |
|---|---|---|
| Length of GI (Fraction of useful data length) | 1/4, 1/8, 1/16, 1/32 | |
| Mapping | QPSK, 16QAM, 64QAM (optionally hierarchical) | |
| Coding | Outer | Reed-Solomon RS(204, 188, t=8) |
| | Inner | Convolutional code with code rate 1/2, 2/3, 3/4, 5/6, 7/8 |
| Interleaver | Outer | Convolutional interleaving |
| | Inner | Bitwise + symbol interleaving |

### B. DTMB System

The TDS-OFDM waveform is selected as the basic transmission scheme for the multicarrier mode of DTMB. The transmitted TDS-OFDM signal can be expressed as:

$$\tilde{x}_{TDS}(i) = \mathbf{I}_{zp} \mathbf{F}_M^H x_M(i) + c_P \quad (5)$$

where $\mathbf{I}_{zp} = [\mathbf{I}_M, \mathbf{0}_{M \times D}]^T$ which pads $D$ rows of zeros to the tail of modulated signals, with $\mathbf{0}$ the null matrix of the dimension given by the subscript. $c_P = [\mathbf{0}_{M \times 1}, c_D]^T$ fills the padded zeros with a preselected PN sequence $c_D$. Other variables are similar to the ones used for the description of DVB-T in the previous paragraph. From (5), after completely removing the PN sequence and its effects due to the channel memory, the received signal can be written in the form of a ZP-OFDM signal. Thus, a straight forward way to demodulate TDS-OFDM signal is to remove the PN sequence and use the demodulation algorithms developed for ZP-OFDM. Several methods are proposed with a variation of performance and complexity in [8]. Within these methods, the overlap-add (OLA) algorithm is the least complex means at the expense of losing channel-irrespective invertibility. The ZP-OFDM-OLA is derived from the fact that by splitting ZP-OFDM signal into upper $M \times 1$ and lower $D \times 1$ part, then padding $M - D$ rows of zeros to the latter part, the same $M \times M$ circulant CIR matrix as in (4) can be manually constructed. This means that ZP-OFDM-OLA has identical equalization method as CP-OFDM.

Fig. 2 shows the block diagram of DTMB system with a PN-subtraction-OLA algorithm. At the receiver side, PN sequence convolved by the channel is first extracted with knowledge of CIR. Then, OLA operation is performed by copying the following GI and adding to the beginning part of an OFDM symbol in order to compensate for the effect of GI on the OFDM symbol due to the channel memory, and to restore the orthogonality between subcarriers. After the OLA processing, ZP-OFDM has roughly the same performance as CP-OFDM [7], [8]. The PN sequences are also used to make channel estimation and synchronization, even if the signal equalization is carried out in the frequency domain like in DVB-T, as evident from Fig. 2.

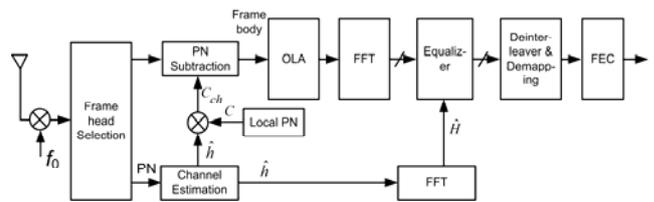

Figure 2. Block diagram of DTMB receiver

Table II presents the main parameters of DTMB by separately listing them in single and multi-carrier mode. It should be mentioned that it is a common "combination" of working modes as shown in [10] and [11], and does not mean that some parameters can only be used in specific mode. Actually, there exists other combinations such as PN595 + C=3780 as shown in [11]. By standard definition, three different PN sequences, PN420, PN945 and PN595, can be used as GI, while the PN 595 has different properties from others in terms of average power, fixed phase and generation method. DTMB supports five types of constellations: 4QAM, 4QAM-NR, 16QAM, 32QAM and 64QAM. The time domain interleaver is the same as the outer interleaver in DVB-T, but with much longer interleaving depth. The interleaving in DTMB is performed over a large number of OFDM blocks to obtain a high diversity gain. The frequency domain interleaving is only used in the multicarrier mode. The following frequency interleaving maps time domain interleaved symbols to 3780 subcarriers in a scrambling order. Concatenated BCH and LDPC are selected as channel coding with three options of code rate. For the sake of comparison fairness, this paper only focuses on the multicarrier mode of DTMB system.

### C. System Comparison

In DVB-T, only the central 83% subcarriers are actually available for data transmission. The remaining FFT points at side parts are deliberately shut down to limit the signal spectrum within 8 MHz analog bandwidth. In order to aid channel estimation and synchronization, continual and scattered pilots are inserted in OFDM symbols, occupying more than 10% subcarriers. There are about 1% subcarriers allocated to the transmission parameter signaling (TPS) which relates to the transmission parameters, e.g. channel coding and modulation. All these factors degrade the spectrum utilization in DVB-T and further introduce a useful data rate loss.

On the other hand, in DTMB, the synchronization and channel estimation are performed by using PN sequences. So there is much less spectrum efficiency loss due to pilots. The only spectrum efficiency degradation comes from the 36 symbols of system information in each 3780-long OFDM block. These symbols take about 1% subcarriers, which is equivalent to the cost of TPS in DVB-T. Eventually, the spectrum utilization of DTMB is about 10% higher than that of DVB-T.

TABLE II. PARAMETERS OF DTMB

|  |  | Single carrier mode | Multicarrier mode |
|---|---|---|---|
| Origin | | Former ADTB-T | Former DMB-T |
| Number of subcarriers | | C = 1 | C = 3780 |
| PN sequence Frame Header | Length | 595 (1/6) | 420 (1/4), 945 (1/9) |
| | Power | Non boost | Boosted by 2 |
| | Phase | Same in a superframe | Different or same |
| Mapping | | 4QAM-NR, 4QAM, 16QAM, 32QAM | 4QAM, 16QAM, 64QAM |
| Interleaver | | Time domain | Time & Frequency domain |
| Coding | Outer | BCH(762, 752) | |
| | Inner | LDPC(7493, 3048), (7493, 4572), (7493, 6096) | |
| | Code rate | 0.4(7488, 3008), 0.6(7488, 4512), 0.8(7488, 6016) | |

GI is also an expense of transmission power and useful data rate. In DVB-T, CP is a duplicate of data part with the same power. However, in DTMB, two types of PN sequence frame header are boosted to obtain better channel estimation and synchronization performance. The boosted PN sequence spends more power than the non-boosted CP given the same GI length.

In order to evaluate the transmission costs in the two systems, all factors mentioned above should be taken into account. An evaluation of power efficiency can be obtained by calculating the ratio of the power allocated to the data subcarriers over all power spent in the transmission. Specifically, in DVB-T, this power efficiency factor can be computed by

$$\gamma_{DVB-T} = \frac{N_{data}}{N_{data} + N_{TPS} + N_{pilot} \times boost} \times \frac{1}{1+GI} \quad (6)$$

where $N_{data}$, $N_{TPS}$, $N_{pilot}$ represent the number of data, TPS and pilot subcarrier, respectively. The *boost* is the boost factor for pilot subcarriers. The *GI* stands for the fraction of GI over data part. In DTMB, a similar expression can be written as:

$$\gamma_{DTMB} = \frac{N_{data}}{N_{data} + N_{info}} \times \frac{1}{1+GI \times boost}. \quad (7)$$

(7) uses the same notation as (6) except that $N_{info}$ represents the number of system information symbols. It should be noticed that this power efficiency factor presents the power allocation in data subcarriers independent of mapping and coding scheme. Or, in other words, it is a measurement of the transmission overhead in terms of power, showing the efficiency of the system data structure. Parameters and the resulting power efficiency factors are presented in Table III. In GI = 1/4 case, although DTMB has higher spectrum utilization ratio of 10%, DTMB and DVB-T have the same power efficiency factor. This is mainly due to the fact that the PN sequence which takes a large portion of transmitted signal, are boosted, decreasing the overall power efficiency. However, this problem is less significant in the GI = 1/9 mode of DTMB. The power efficiency factor in this case is 0.81, which is not only significantly higher than the equivalent GI = 1/8 mode in DVB-T, but is even slightly higher than it in the case of the minimum GI of 1/32.

TABLE III. PARAMETER COMPARISON BETWEEN DVB-T AND DTMB

|  | DVB-T | | DTMB |
|---|---|---|---|
|  | *2K mode* | *8K mode* | *C = 3780* |
| FFT size | 2 048 | 8 192 | 3 780 |
| Nb. of subcarriers | 1 705 | 6 817 | 3 780 |
| Nb. of data subcarriers | 1 512 | 6 048 | 3 744 |
| Subcarrier spacing | 4 464 Hz | 1 116 Hz | 2 000 Hz |
| Signal bandwith | 7.61 MHz | 7.61 MHz | 7.56 MHz |
| OFDM symbol duration | 224 μs | 896 μs | 500 μs |
| Power efficiency factor | 0.66 (GI=1/4), 0.73 (GI=1/8), 0.77 (GI=1/16), 0.79 (GI=1/32) | | 0.66 (GI=1/4), 0.81 (GI=1/9) |

Besides increasing the spectrum utilization, PN sequence makes it possible to achieve a faster channel acquisition in DTMB. In the DVB-T system, complete channel estimation can be performed by using pilots from four consecutive OFDM blocks, while, in DTMB, it can be made for every block relying on its own PN sequence. This feature is expected to make DTMB more robust in high mobility scenario.

III. SIMULATION RESULTS

In this section, we analyze the bit error rate (BER) performance of the two systems in additive white Gaussian noise (AWGN), Ricean ($F_1$) and Rayleigh ($P_1$) channels. The latter two channels are specified in [1]. The length of GI is set to 1/4, because it is the only common option in both systems. Using the same ratio of GI will introduce identical performance loss due to the time domain redundancy which plays an important role in combating multipath effects. Another way to guarantee the fairness of comparison is to find a pair of working modes with approximately the same useful bitrate. In Table IV, three working modes are picked from each system, representing low, medium and high throughput applications,

TABLE IV. SIMULATION PARAMETERS AND USEFUL BIT RATES AT GI =1/4

| Mode | System | Mapping | Code Rate | | Bitrate (Mbps) | |
|---|---|---|---|---|---|---|
| | | | *No outer code* | *With outer code* | *No outer code* | *With outer code* |
| 1 | DVB-T | QPSK | 1/2 | 0.46 | 5.4 | 4.98 |
| 2 | | 16QAM | 3/4 | 0.69 | 16.2 | 14.93 |
| 3 | | 64QAM | 3/4 | 0.69 | 24.3 | 22.39 |
| 4 | DTMB | QPSK | ≈ 0.4 | ≈ 0.4 | 4.88 | 4.81 |
| 5 | | 16QAM | ≈ 0.6 | ≈ 0.6 | 14.63 | 14.44 |
| 6 | | 64QAM | ≈ 0.6 | ≈ 0.6 | 21.96 | 21.66 |

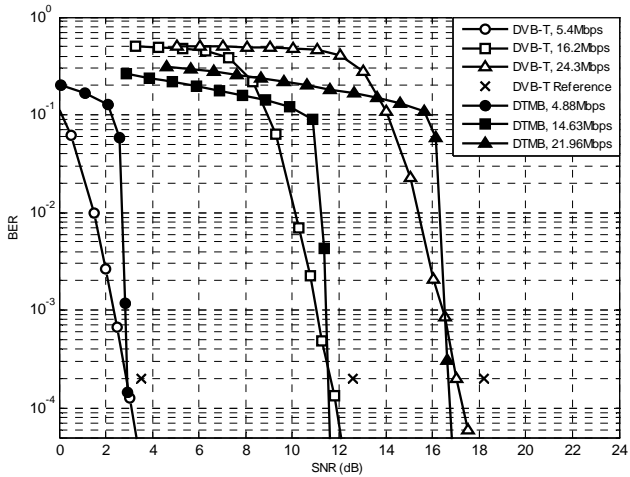

Figure 3. BER comparison of DVB-T and DTMB in AWGN channel without outer code, GI = 1/4

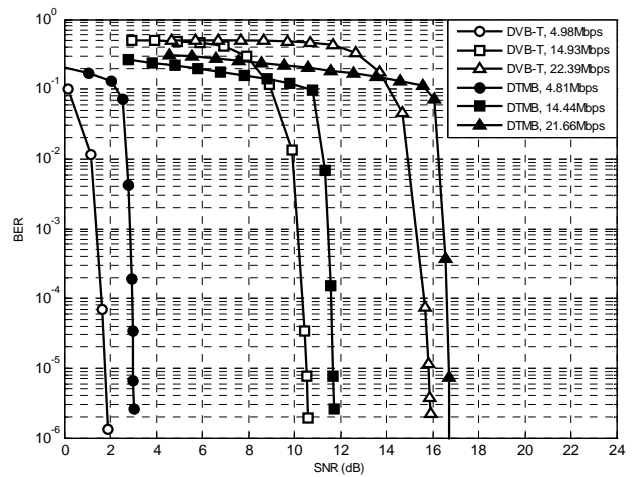

Figure 4. BER comparison of DVB-T and DTMB in AWGN channel with outer code, GI = 1/4

respectively. Corresponding useful bitrates are also given.

The 2K mode is selected as the representative of DVB-T. In DTMB, the interleaving depth is chosen as 240. The LDPC decoder adopts the message-passing algorithm with a maximum iteration times of 50 which is a good trade-off between error correcting performance and time consuming.

All the simulation results are presented in terms of BER versus signal to noise ratio (SNR) which is defined by the average signal power over noise power.

### A. In AWGN Channel

Fig. 3 gives the comparison results without taking into account the outer code in AWGN channel. In DVB-T, for quasi-error-free (QEF) reception, the projected post-RS BER is less than $10^{-11}$, requiring a post-Viterbi BER of less than $2 \times 10^{-4}$ which is taken for evaluation here. C/N references given by [1] are also printed as cross in the figure. One has to be careful about the fact that these C/N references actually correspond to the ratio of the power on data subcarriers over the power of noise, *without* considering the power of pilots and GI as well as the inactive subcarriers. However, the SNR used here is the exact ratio of the received power *including* the power spent for pilot subcarriers over the power of noise. For this reason, the averaged SNR is slightly smaller than C/N. Specifically in DVB-T, the references given in C/N should be shifted 0.46 dB to the left to get the corresponding references in SNR. After this shift, it can be observed that simulated curves of mode 1 to 3 exactly pass over the references, proving the correctness of our simulations. The BER performance of DTMB without outer code is also shown in the same figure. It can be seen that in terms of SNR, the two systems have almost the same performance in low and medium data rate cases, while in the high data rate situation, DTMB is 0.4 dB better when BER equals to $2 \times 10^{-4}$.

Fig. 4 shows the simulation results with outer code in both systems. Profiting from the RS (204, 118) with 8 byte error correction capability and the interleaving between inner and outer code, the performance of DVB-T is significantly improved, exhibiting a sharp flop. However, the BCH (762, 752) code of the DTMB system can only correct one bit error and does not exhibit any effect at BER level of $10^{-4}$. No improvement can be observed when the outer code is added to the simulation in DTMB. From Fig. 4, DVB-T is 1.2 dB, 1.1 dB and 1.0 dB better than DTMB at BER=$5 \times 10^{-5}$ in three modes.

Comparing Fig. 3 and Fig. 4, we can see different philosophies for the two systems. In DVB-T, the task of error correction is shared by inner and outer coding. So each of them should be sufficiently effective and the interleaving between them is necessary. On the other hand, in DTMB, the duty of forward error correcting is mainly fulfilled by the LDPC code. The LDPC code has such superior performance that the major role of BCH is actually to adapt the data frame lengths [11]. So we can understand why the interleaving process between inner and outer code are omitted in DTMB. It is more reasonable to compare the two systems with full error correcting ability. By understanding this, reset comparisons are carried out only in the "with outer code" case.

### B. In Multipath Channels

Fig. 5 and Fig. 6 give the simulation results in Rayleigh channel and Ricean channel, respectively. The BER is measured at the output of outer decoder for each system. All simulations are carried out under the assumption of perfect channel estimation and synchronization. In the $P_1$ channel, DVB-T is 0.7 dB better than DTMB at BER = $5 \times 10^{-5}$ in the low throughput case, and this difference decreases at lower BER level. Moreover, DTMB outperforms DVB-T in the other two cases at the same BER and a greater difference can be foreseen at even lower BER level.

In the $F_1$ Ricean channel, the presence of a line of sight path makes the performance close to that obtained in AWGN. Hence, a similar conclusion can be made about the results in $F_1$ as in AWGN, though some degradations of DVB-T can be noticed at higher code rates and constellations.

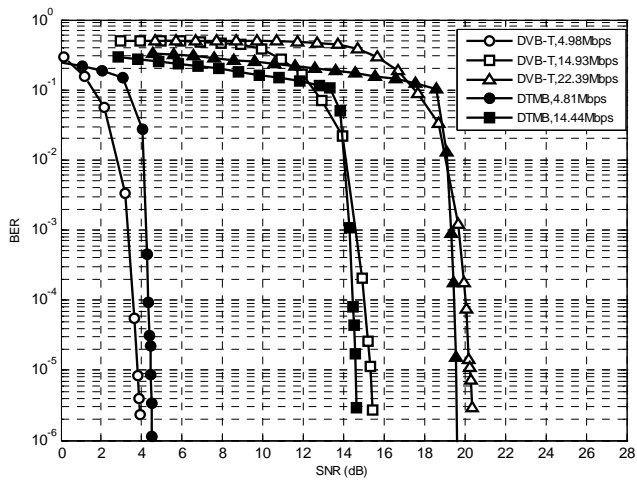

Figure 5.  BER comparison of DVB-T and DTMB in $P_1$ channel with outer code, GI = 1/4

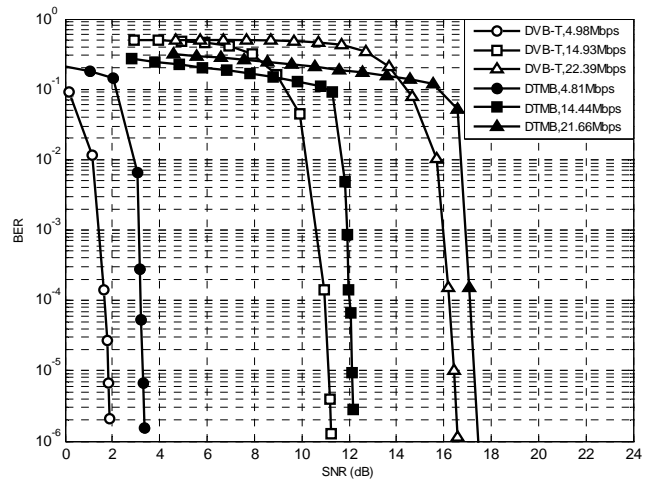

Figure 6.  BER comparison of DVB-T and DTMB in $F_1$ channel with outer code, GI = 1/4

From Fig. 5, Fig. 6 and comparing with the results in AWGN, the performance degradation in $P_1$ and $F_1$ channels is stronger in DVB-T. This can be understood by the fact that DTMB exhibits a better adaptability to the multipath channel than DVB-T, thanks to the LDPC code combined with the extremely deep interleaving.

## IV. CONCLUSION

In this paper, two DTTB standards － DVB-T and DTMB, based on different GI approaches, are presented, compared and analyzed. Discussions on power utilization are taken to analyze the transmission overhead cost in power perspective. Power efficiency factors demonstrate that DTMB has equivalent power efficiency as DVB-T in long GI (1/4) case, while the short GI (1/9) case of DTMB has a better power efficiency compared with all cases in DVB-T. Simulation results show that the performances of two systems are really close. When the GI equals to 1/4, DVB-T seems to outperform DTMB in terms of BER versus SNR in AWGN and $F_1$ channels, while DTMB enjoys better performance in $P_1$ channel because its channel coding and interleaving scheme is more effective in strong fading environments. In further studies, the comparison of the systems will be carried out in time selective fading channels implementing real channel estimation algorithms.

REFERENCES

[1] Digital video broadcasting (DVB); Framing structure, channel coding and modulation for terrestrial television, European Standard (EN) 300 744 V1.5.1, European Telecommunications Standards Institute (ETSI), Nov. 2004.

[2] Advanced Television Systems Committee, "ATSC Digital Television Standard," Document A/53, Sept. 1995.

[3] Association of Radio Industries and Businesses, "Terrestrial Integrated Services Digital Broadcasting (ISDB-T) — Specifications of Channel Coding, Framing Structure, and Modulation," Sept. 1998.

[4] Framing Structure, Channel Coding and Modulation for Digital Television Terrestrial Broadcasting System, Chinese National Standard GB 20600-2006.

[5] J. Wang, Z. Yang, C. Pan, M. Han, L. Yang, " A combined code acquisition and symbol timing recovery method for TDS-OFDM," Broadcasting , IEEE Transactions on, vol. 49, pp. 304-308, Sept. 2003.

[6] M. Muck, M. de Courville, P. Duhamel, "A pseudorandom postfix OFDM modulator — semi-blind channel estimation and equalization," Signal Processing, IEEE Transactions on, vol.54, no.3, pp. 1005-1017, March 2006.

[7] H. Steendam, M. Moeneclaey, "Different Guard Interval Techniques for OFDM: Performance Comparison," in Proc. Of MC-SS'07, Herrsching, Germany, May 7-9, 2007.

[8] B. Muquet, Z. Wang, G.B. Giannakis, M. de Courville, P. Duhamel, "Cyclic prefixing or zero padding for wireless multicarrier transmissions?," Communications, IEEE Transactions on, vol.50, no.12, pp. 2136-2148, Dec 2002.

[9] A. Stephan, E. Guéguen, M. Crussière, J.-Y. Baudais, and J.-F. Hélard, "Optimization of Linear Precoded OFDM for High-Data-Rate UWB Systems," Wireless Communications and Networking, EURASIP Journal on, vol. 2008, Article ID 317257, 2008.

[10] J. Song, Z. Yang, L. Yang, K. Gong, C. Pan, J. Wang, Y. Wu, "Technical review on Chinese Digital Terrestrial Television Broadcasting Standard and Measurements on Some Working Modes," Broadcasting, IEEE Transactions on , vol.53, no.1, pp.1-7, March 2007.

[11] W. Zhang, Y. Guan, W. Liang, D. He, F. Ju, J. Sun, "An Introduction of the Chinese DTTB Standard and Analysis of the PN595 Working Modes, " Broadcasting, IEEE Transactions on , vol.53, no.1, pp.8-13, March 2007.